\documentclass[12pt]{article}
\usepackage{amsfonts}
\usepackage{amsmath}
\usepackage{amssymb}
\usepackage{array}
\usepackage{bigints}
\usepackage{booktabs}
\usepackage[nosort]{cite}
\usepackage{color}
\usepackage{dsfont}
\usepackage{float}   
\usepackage{framed}
\usepackage{graphicx}
\usepackage{indentfirst}
\usepackage{mathrsfs} 
\usepackage{multirow}
\usepackage{pdflscape}
\usepackage{setspace}
\usepackage{subdepth}
\usepackage{subfig}
\usepackage{titlesec}
\usepackage[dotinlabels]{titletoc}
\usepackage{wrapfig}
\usepackage[all]{xy}
\usepackage{young}
\usepackage[vcentermath]{youngtab}
\usepackage{relsize}
\usepackage{stackengine}
\usepackage{datetime}

\usepackage{hyperref}

\numberwithin{equation}{section}

\usepackage{verbatim}

\newcommand{\be}{\begin{equation}}
\newcommand{\ee}{\end{equation}}

\def\({\left(} \def\){\right)}
\def\[{\left[} \def\]{\right]}

\def\sgn{\text{sgn}}

\def\mO{\mathcal{O}}

\def\mI{\mathcal{I}}

\def\eps{\epsilon}

\def\vphi{\varphi}

\newcommand{\bea}{\begin{eqnarray}}
\newcommand{\eea}{\end{eqnarray}}

\newcommand{\bml}{\begin{multline}}
\newcommand{\emll}{\end{multline}}

\usepackage[left=2.5cm,right=2.5cm,top=2.5cm,bottom=2.5cm]{geometry}
\titleformat{\section}{\normalfont\bfseries}{\thesection.}{4pt}{}
\titlespacing{\section}{0pt}{22pt}{6pt}

\titleformat{\subsection}{\normalfont\itshape}{\thesubsection.}{4pt}{}
\titlespacing{\subsection}{0pt}{18pt}{6pt}

\titleformat{\subsubsection}{\normalfont\itshape}{\thesubsubsection.}{4pt}{}
\titlespacing{\subsubsection}{0pt}{16pt}{6pt}

\def\ie{\begin{equation}\begin{aligned}}
\def\fe{\end{aligned}\end{equation}}

\def\bar{\overline}

\def\1{{\mathds 1}}

\DeclareFontShape{OT1}{cmr}{mx}{n}%
    {<->cmr10}{}
\newcommand{\mytitlefont}{\fontseries{mx}\selectfont}
\DeclareMathAlphabet{\titlemath}{OT1}{cmr}{mx}{n}

\begin{document}


\begin{titlepage}

\begin{center}

~\\[2cm]

{\fontsize{20pt}{0pt} \mytitlefont A line of CFTs:    from generalized free fields to SYK}

~\\[0.5cm]

{\fontsize{14pt}{0pt} David J.~Gross and Vladimir Rosenhaus}

~\\[0.1cm]

\it{Kavli Institute for Theoretical Physics}\\ \it{University of California, Santa Barbara, CA 93106}

~\\[0.8cm]

\end{center}

\noindent  We point out that there is a simple variant of the SYK model, which we call cSYK, that is $SL(2,R)$ invariant for all values of the  coupling. The modification consists of replacing the UV part of the SYK action with a quadratic bilocal term. The corresponding bulk dual is a non-gravitational theory in a rigid AdS$_2$ background. At weak coupling cSYK is a generalized free field theory; at strong coupling, it approaches the infrared of SYK. The existence of this line of fixed points explains the previously found connection between the three-point function of bilinears  in these two theories  at large $q$.

\vfill 


\end{titlepage}

\vfill


\tableofcontents

~\\[.1cm]

\section{Introduction}
The simplest large-$N$ conformal field theory consists of $N$  massless free fields. This theory is $O(N)$ symmetric, and the holographic dual of the singlet sector is Vasiliev higher spin theory \cite{Fradkin:1987ks, Vasiliev:1999ba, Klebanov:2002ja}. Although simple on the boundary, the bulk theory is only partially understood, either as a quantum field theory or as some kind of string theory \cite{Chang:2012kt, Gaberdiel:2014cha}. Part of the difficulty in understanding the bulk theory is the infinite-dimensional  gauge symmetry in the bulk that arises from the infinite number of local conserved currents on the boundary, and the complexity of the massless higher spin bulk theories.  Also, the free $O(N)$ model, and  most other examples of  large $N$ dualities, lack the marginal coupling of supersymmetric gauge theory that allows one to continuously go from weak to strong coupling.

An even simpler setting, and one that has been less explored, is to consider the free $O(N)$ model in $0+1$ dimensions. Taking the fields to be Majorana fermions, the action is, 
\be \label{top}
S_{top} = \sum_{i =1}^N \int d\tau\, \chi_i \, \partial_{\tau} \chi_i~.
\ee 
Unlike its higher dimensional cousins, this theory is topological: the action is invariant under arbitrary time reparametrizations and consequently the Hamiltonian is zero. The bulk dual is presumably some topological cousin of Vasiliev theory for AdS$_2$ \cite{Bengtsson:1986zm, Fradkin:1989uh,Vasiliev:1995sv, Alkalaev:2013fsa, Grumiller:2013swa, Alkalaev:2014qpa, Mezei:2017kmw}.  However, for addressing the questions  mentioned above, and for finding some hypothetical new theory of extended objects, one needs a  dynamical model.
 A simple deformation of the topological free theory that preserves one-dimensional conformal invariance ($SL(2,R)$), while breaking the infinite-dimensional time reparameterization symmetry,  is a non-local theory with action, 
\be  \label{top1}
S_0 =-\Delta\sum_{i =1}^N\int d\tau_1 d\tau_2\, \chi_i(\tau_1)\, \frac{\sgn(\tau_{1}-\tau_{2})}{|\tau_{1}-\tau_{2}|^{2 - 2\Delta}}\, \chi_i(\tau_2) ~.
\ee
When $\Delta$ goes to zero, this reduces to the topological theory of free Majorana fermions, but otherwise is a generalized free theory, in which the dimension of the fermion has been shifted from zero to  $\Delta$. One can view this action as the coupling of an operator $\chi_i$, with dimension $\Delta$,  to its shadow operator of dimension $1- \Delta$ \cite{Ferrara:1972xe, Ferrara:1972uq}. The advantage of  this theory is the absence of gauge symmetry in the bulk dual and the ability, in AdS$_2$,  to  take all the bulk  fields, that would ordinarily be massless, higher spin fields, to simply be massive scalars. The bulk dual of this theory does not have gravity; it is simply a theory of an (infinite) number of scalars on a fixed AdS$_2$ background. For studying gravitational questions, this model is not obviously helpful. However, it  may be useful for the purposes of constructing a dual bulk theory of extended objects.

 This generalized free theory  has another advantage, which is in fact how we were lead to it: one can consider adding a local deformation by the interaction term of the SYK model \cite{Kitaev, SY}, 
\be
S^{ \rm int}_{SYK} =\frac{(i)^{\frac{q}{2}}}{q!}\sum_{i_1, \ldots, i_q = 1}^N \int d\tau \, J_{i_1\, i_2\, \ldots i_q}\, \chi_{i_1}\chi_{i_2}\, \cdots \chi_{i_q}~,
\ee
where $J_{i_1, i_2, \ldots i_q}$ is chosen from a Gaussian ensemble with variance proportional to $J^2/N^{q-1}$. 
If we choose $\Delta=1/q$, in (\ref{top1}), then this deformation is marginal. Furthermore, the theory $S_0 + S^{\rm int}_{SYK}$ is classically  $SL(2,R)$ invariant for all values of the coupling $J$, which now is a dimensionless constant. We will refer to this theory as conformal SYK (cSYK). We shall argue in the following that there are no quantum anomalies, at least to leading order in $1/N$, and thus cSYK is conformally invariant for all $J$; in other words, we have a line of fixed points. A theory with a line of fixed  points is  rare; a notable case is maximally supersymmetric $\mathcal{N}=4$ Yang-Mills: at  strong 't Hooft coupling it is  dual to string theory with a small string length, and at weak 't Hooft coupling to string theory with large string length. The existence of a line of fixed points is highly nontrivial and, in the case of $\mathcal{N}=4$ Yang-Mills,  is due to supersymmetry. On the other hand, for cSYK the line of fixed points is less obvious. It appears to be a consequence of the non-locality and the anti-commuting fields. 

For large $J$, the interacting part of the action, $S_{SYK}^{\rm int}$, dominates over the free part, either $S_{top}$ or $S_0$, for SYK or cSYK, respectively, and the behavior of cSYK approaches that of SYK. All the results for the infrared (large $J$) of SYK, in particular the dimensions of the $O(N)$ invariant bilinear singlets \cite{Kitaev, PR, MS} and their three-point functions  \cite{GR2}, can be easily generalized to cSYK for any $J$, thus yielding the masses $m_n$ and cubic couplings $\lambda_{n m k}$ for bulk fields $\phi_n$ dual to the bilinear singlets $\mO_n$, for any value of $J$. 

The three-point functions of the bilinear singlets, at the infrared fixed point of SYK, were studied in \cite{GR2}. There were two classes of Feynman diagrams that contributed, which we denoted by planar and ``contact'' diagrams. It was observed that the contribution of the planar diagrams, for large $q$,  is related in a simple way to the three-point function of bilinears in the generalized free theory (\ref{top1}). Having a line of fixed points interpolating between these two theories allows us to understand this relation.

The paper is organized as follows. In Sec.~\ref{sec:bi} we show how bilocal actions can be obtained from local actions wherein the $\chi_i$'s couple to  towers of auxiliary fields which are then integrated out. This provides a ``physical'' setting for such a non-local action, at the price of introducing an infinite number of new degrees of freedom. This is a familiar fact; in the current context it makes it clear that a CFT$_1$ of this type should be viewed as a subsector of a 2d theory,  a necessary condition for a hypothetical duality with a non-gravitational AdS$_2$ theory.  In Sec.~\ref{sec:line} write down the two-point and three-point functions of the $O(N)$ invariant bilinear singlets $\mO_n$ in cSYK. One distinction between cSYK and the infrared of SYK are the correlation functions involving the lowest dimension singlet, $\mO_0$. In SYK these break conformal invariance; in cSYK these preserve conformal invariance. On the bulk side, this translates into the statement that the gravitational sector of the dual of SYK is Jackiw-Teitelboim gravity, whose infrared behavior is dominated by the reparameterization  fluctuations of the AdS boundary; whereas the dual of cSYK is a  field theory on a fixed AdS background and the dual of $\mO_0$ is just the lightest bulk field. We make some brief remarks on this in Sec.~\ref{sec:bulk}. In appendix~\ref{ap:three} we record the three-point functions of the $\mO_n$ for cSYK, comparing weak and strong coupling.

\section{Bilocal Action } \label{sec:bi}
The action that we will be studying, (\ref{top1}), is bilocal in time. As such, it is difficult to make sense of it on its own, and it is unclear if it is physical.~\footnote{Actions that are instead bilocal in space, as for instance the long-range Ising model recently discussed in \cite{Behan:2017emf}, are clearly physical. } One could focus on (\ref{top1}) in the small $\Delta$ limit, regarding $\Delta$ as a regulator away from the local action (\ref{top}), but it is better to understand the action in general. Bilocal actions of this type are familiar in a number of contexts, and should be viewed as arising  from some local action after integrating out auxiliary degrees of freedom. 

\subsection*{Bulk Scalar}
As an illustration, consider a massive scalar field in AdS$_{d+1}$, 
\be \label{Ifree}
I_{free} = \frac{1}{2} \int d^{d+1} x \sqrt{g} \[ (\partial \phi)^2 + m^2 \phi^2\]~.
\ee
As is very familiar, integrating out the bulk degrees of freedom, which can be done exactly as  (\ref{Ifree}) is quadratic, gives rise to a bilocal action with a kernel that is a conformal two-point function with dimension set by the mass. In particular, working in Euclidean coordinates, 
\be
d s^2 = \frac{1}{z^2} \( d x^2 + d z^2\)~,
\ee
and letting $\phi(z, x) \rightarrow z^{d-\Delta_0} \phi_0(x)$ as $z\rightarrow 0$, where $\Delta_0 (\Delta_0 - d) = m^2 $, one can solve for $\phi(z, x)$ in terms of $\phi_0$, 
\be \label{phiK}
\phi(z, x) = \int d^d x'\, K_{\Delta_0}(z, x\, |\, x')\, \phi_0(x')~, \ \ \ \ \ K_{\Delta_0}(z, x \, |\, x') = \frac{\pi^{-d/2} \Gamma(\Delta_0)}{\Gamma( \Delta_0 - d/2)} \frac{ z^{\Delta_0}}{\(z^2 + (x - x')^2\)^{\Delta_0}}~.
\ee
Plugging $\phi$ into the action (\ref{Ifree}), the action just becomes a boundary term \cite{Witten:1998qj}, 
\be \label{Inonl}
I = - (\Delta_0 - d/2) \pi^{-d/2} \frac{\Gamma(\Delta_0)}{\Gamma(\Delta_0 - d/2)} \int d^d x\, d^d x'\, \frac{\phi_0(x) \phi_0(x')}{|x - x'|^{2 \Delta_0}}~.
\ee

An equivalent viewpoint, which will be more useful for us, is to  think of the system as consisting of the free scalar $\phi(z, x)$ for $z\geq0$, along with another field $\phi_0(x)$ that lives on the boundary and is linearly coupled to $\phi(z=0, x)$. (So  $\phi_0(x)$ can be thought of a source for $\phi$). The action of the full system 
\be \label{In2}
I = I_{free}+  I_{source}
\ee
where, 
\be \label{Isource}
I_{source} = \alpha\, \epsilon^{d+1- \Delta_0}\int d^{d+1} x \sqrt{g}\, \delta( z - \epsilon)\,  \phi(z, x) \phi_0(x)~,
\ee
where $\alpha$ is a constant, and we have regulated the boundary by moving it inward to $z=\epsilon$. For the purposes of this action, the allowed solutions are those that decay as $z^{\Delta}$ for small $z$. The reason this is the same  as the previous action (\ref{Inonl}) is the following: if we just think of the interior of AdS, and place sources somewhere, then the field is given by the convolution of the sources with the bulk two-point function $G_{bulk}(z, x\, |\, z', x')$ (which is constructed out of the normalizable modes). In the limit that $z'\rightarrow 0$, 
\be
G_{bulk}(z, x\, |\, z', x') \rightarrow \frac{1}{2\Delta_0 - d} z'^{\Delta_0} K_{\Delta_0}(z, x\, |\, x')~,  \ \ \ \ \ \ \text{as  } z'\rightarrow 0~,
\ee
so (\ref{In2}) implies (\ref{phiK}), and in particular that $\alpha = 2\Delta_0 - d$.
The action (\ref{Inonl}) is the kind of bilocal action we are seeking; it is the action for a generalized free field $\phi_0$ of dimension $\Delta = d-\Delta_0$. With standard quantization, as we have discussed, one can achieve $\Delta_0 \geq d/2$, where the lower bound is set by the BF bound on the mass. With alternate quantization \cite{Klebanov:1999tb} this can be extended to $\Delta_0\geq d/2 - 1$.

\subsection{A Tower of Auxiliary Fields} \label{sec:tower}
For generalizing (\ref{In2}) it is useful to rewrite it as an inherently $d$ dimensional theory. 
We decompose the field in terms of its radial eigenfunctions, which are Bessel functions, 
\be
\phi(z, t, x) = \int_0^{\infty} d\lambda \, z^{d/2}\, J_{\nu} (\lambda z)\,  \sqrt{\lambda}\, \vphi_{\lambda}(t,x)~,
\ee
where $\Delta_0 = d/2 + \nu$, with $\nu = \sqrt{m^2 + d^2/4}$. Inserting this expansion of $\phi(z,t,x)$ into the action (\ref{In2}) and restoring the AdS scale $L$, we get, 
\be \label{Itower2p5}
I = \int d^{d} x\int_0^{\infty} d\lambda \[ \frac{1}{2} \( (\partial \vphi_{\lambda})^2 + \lambda^2 L^{-2} \vphi_{\lambda}^2\) + L^{-\nu - 1}\, \bar{\alpha}\, \lambda^{\nu+\frac{1}{2}}\, \vphi_\lambda \phi_0 \]~,
\ee
where we have defined $\bar{\alpha} =2^{1-\nu}/\Gamma(\nu)$. So we have a tower of fields $\vphi_{\lambda}$, of mass $ \lambda^2 L^{-2}$,  all of which are linearly coupled to $\phi_0$. Let us integrate out $\vphi_{\lambda}$. This gives an effective action, 
\be \label{Ieff}
I_{eff} = \frac{1}{2}\int d^d x d^d x'\, \phi_0(x) K(x, x') \, \phi_0(x')~,
\ee
where, 
\be \label{K1}
K(x, x') =   \bar{\alpha}^2 \int_0^{\infty} d \lambda \, {\lambda}^{2 \nu+1}\,  \langle x| \frac{1}{\Box - {\lambda}^2} |x'\rangle~,
\ee
and so reproduces (\ref{Inonl}), wherein the non-locality arises since we have integrated out massless fields.

\subsection*{General and Finite Temperature} \label{sec:general}
To obtain the appropriate action at finite temperature, we could repeat the procedure, considering an AdS-Schwarzschild background rather than the Poincare patch.~\footnote{ Note that in AdS$_2$ one get between different backgrounds through a change of coordinates. However, if one chooses to maintain Poincare patch coordinates, then the necessary source will no longer be at constant $z=\epsilon$, but rather along some trajectory $z(t)$ that one can find through a change of coordinates to AdS-Schwarzschild coordinates, with the radial coordinate there set to $\epsilon$. } 
This would be more involved, and is in fact unnecessary, since a clear generalization of (\ref{Itower2p5}) is, 
\be \label{ItowerGen}
I = \int d^{d} x\int_0^{\infty} d\lambda \[ \frac{1}{2} \( (\partial \vphi_{\lambda})^2 + \lambda^2 L^{-2} \vphi_{\lambda}^2\) + j(\lambda) \vphi_\lambda \phi_0 \]~,
\ee
for some $j(\lambda)$ that can depend on $L$ and other scales one may choose to introduce. Integrating out $\vphi_{\lambda}$ leads to the bilocal action (\ref{Ieff}) with the kernel, 
\be \label{K2}
K(x, x') =  \int_0^{\infty} d \lambda \, j(\lambda)^2\, \langle x| \frac{1}{\Box -\frac{\lambda^2}{L^2}} |x'\rangle~.
\ee

In particular, it is now easy to find the necessary coupling $j(\lambda)$ to achieve the finite temperature generalized free field action for a boson in $0+1$ dimension.  In other words, we would like to find a $j(\lambda)$ such that the kernel (\ref{K2}) is the inverse of the finite temperature two-point function, $G_{\Delta}(\omega_n)$, for a conformal scalar field of dimension $\Delta$,
\be \label{KiG}
\!\!\! K(\omega_n) =-\!\frac{2\pi \tan \pi \Delta}{(2\Delta-1)}\frac{1}{G_{\Delta}(\omega_n)}~, \ \ \ \ \ \ \ \  G_{\Delta}(\omega_n) = \(\frac{2\pi}{\beta}\)^{2\Delta-1}\!\!\! \frac{\pi}{\cos \pi \Delta\, \Gamma(2\Delta)}\frac{\Gamma\(\Delta + \frac{\beta \omega_n}{2\pi}\)}{\Gamma\( 1- \Delta + \frac{\beta \omega_n}{2\pi}\)}~,
\ee
where $\omega_n = 2\pi n/\beta$ are the Matsubara frequencies. One can see that $G_{\Delta}(\omega_n)$ satisfies the property, 
\be
 \frac{1}{G_{\Delta}(\omega_n)} = \frac{(2\Delta_0-1)}{2\pi \tan \pi \Delta_0}\, G_{\Delta_0}(\omega_n)~, 
  \ee
 where we  have defined $\Delta_0 = 1-\Delta$. From (\ref{K2}) we have that, 
 \be
 K(\omega_n) = - \int_0^{\infty} d\lambda \frac{j(\lambda)^2}{\omega_n^2 + \frac{\lambda^2}{L^2}}~.
 \ee
Therefore, we can achieve the desired (\ref{KiG}) with the choice of coupling,
 \be \nonumber
 j(\lambda)^2 =\!\frac{ 2}{L}\frac{\lambda}{L}\, \rho_{\Delta_0}(\lambda/L) =\!\frac{ 2}{L}\frac{\lambda}{L} \frac{1}{\pi \Gamma(2\Delta_0)}\(\frac{2\pi}{\beta}\)^{2\Delta_0-1}\!\! \sinh\( \frac{\beta \lambda}{2 L}\) \Gamma\(\Delta_0 - \frac{i \lambda \beta}{2\pi L}\)\Gamma\(\Delta_0 + \frac{i \lambda \beta}{2\pi L}\) ~,
 \ee
 where we have made use of the spectral density $\rho_{\Delta_0} (\lambda)$ for a finite-temperature conformal scalar of dimension $\Delta_0$ \cite{PGKS}. The dimension of $j(\lambda)^2$ is $2\Delta_0+1$. Since the dimension of $\vphi_{\lambda}$ is $-1/2$ and the dimension of $\phi_0$ is $1-\Delta_0$, the interaction $j(\lambda) \vphi(\lambda) \phi_0$ in (\ref{ItowerGen}) is dimension $1$, as it should be.

One may find it strange that we must change the theory (\ref{ItowerGen}) in order to change the temperature of the effective theory for $\phi_0$. In fact, this is appropriate. Once we integrate out the $\vphi_{\lambda}$ degrees of freedom, we no longer have the ability to excite them, so changing the state of the $\vphi_{\lambda}$ leads to a different effective theory for the $\phi_0$. 

In $0+1$ dimensions, the action (\ref{ItowerGen}) is reminiscent of the Caldeira-Leggett model \cite{CL}. There one has a harmonic oscillator $\phi_0$ coupled to a bath of oscillators $\vphi_{\lambda}$. Tracing out the bath gives a Fokker-Planck equation for the reduced density matrix of the $\phi_0$ oscillator. The Heisenberg equations of motion for $\phi_0$ are given by the generalized Langevin equations, which are nonlocal in time, containing a friction term $\int dt' \gamma(t-t') \dot{\phi_0}(t')$.  With an appropriate choice of the coupling $j(\lambda)$, such as the one  Caldeira and Leggett picked, $\gamma(t-t')$ can be made local, $\gamma \delta(t-t')$, resulting in a standard dissipation term. The difference between the Caldeira-Leggett setup and ours is that we are integrating out the $\vphi_{\lambda}$ degrees of freedom rather than tracing them out; we have preformed an exact rewriting of the theory, as is standard in, for instance, RG analysis. 

 Our setup is perhaps more similar to the Kondo problem, in which an impurity is coupled to a bath. Indeed, in that context one can employ dynamical mean field theory to obtain an effective action for the impurity, which is of the form of the bilocal action (\ref{Ieff}), with the kernel fixed self-consistently \cite{DMFT}.

\subsection*{Free Energy and Entropy}
The advantage of rewriting the nonlocal action in terms of a local action is that we now have a notion of energy, and a notion of a Hilbert space. This allows us to study the entropy of the theory. 

Let us consider the partition function for our general action (\ref{ItowerGen}),
\be
Z= \int D\vphi_{\lambda} D\phi_0\, e^{-I} = Z_{free} Z_{eff}~,
\ee
where we have noticed that it factorizes (after an appropriate change of variables $\vphi_{\lambda}$), into a partition function  $Z_{free}$ for the free part of the action, 
\be \label{IFREE}
I_{free} =   \frac{1}{2}\int d^{d} x\int_0^{\infty} d\lambda \( (\partial \vphi_{\lambda})^2 + \lambda^2 L^{-2} \vphi_{\lambda}^2\) ~,
\ee
and a partition function $Z_{eff}$ for the effective action for $\phi_0$ given by (\ref{Ieff}) with kernel (\ref{K2}). The entropy of the full system, involving the $\vphi_{\lambda}$ fields and $\phi_0$ fields, is thus given by a sum of entropies coming from the partition functions of these two pieces, respectively. 

The entropy coming from the free action (\ref{Ieff}) is simply, $\frac{1}{2} \sum_{n} \log |K(\omega_n)|$, up to a constant.
This is in $0+1$ dimensions; if there are more dimensions, then there is also a momentum integral.
In $0+1$ dimensions, this entropy is temperature independent. Furthermore, if $K$ is the kernel appropriate for a generalized, fermionic, conformal free field at finite temperature, (\ref{KiG}), then this entropy is just the ground state of entropy of SYK \cite{GPS, Kitaev}.

Next, let us look at the free energy resulting  from the action  (\ref{IFREE}). Placing  the system in a box of volume $V_{d-1}$, we get,
\be \label{Ffree2}
F_{free} = \beta^{-1} V_{d-1} \int_0^{\infty} d \lambda\, \int \frac{d^{d-\! 1} p}{(2\pi)^{d-1}}\, \log\( 2 \sinh\frac{\beta}{2} \sqrt{\lambda^2/L^2 + p^2}\)~.
\ee
This is proportional to the free energy of gas of massless bosons in $d+1$ spacetime dimensions, $F_{d+1}^{gas}$. Specifically, $F_{free}/V_{d-1}= \pi\,L\,  F_{d+1}^{gas}/V_{d}~$.
By having an infinite tower of fields, we are imitating an extra dimension. 
From the free energy we get that the entropy scales as,
\be
S_{free} \propto (d+1)\frac{ V_{d-1} L }{\beta^d}~.
\ee

So, for $d=1$, we have a one-dimensional system, (\ref{ItowerGen}), that is not a CFT and whose entropy has two pieces: a contribution that scales with the temperature, and a contribution that is constant. We can isolate the part of the system that gives the ground state entropy by integrating out the $\vphi_{\lambda}$ degrees of freedom. This gives us a CFT$_1$, with an entropy that is temperature independent.

 It has been argued that a CFT$_1$ can only be topological, because in one spacetime dimension, by dimensional analysis, the entropy must be a constant \cite{Jensen:2011su}. We have obtained a  CFT$_1$ with dynamics. 
Perhaps one can say that we have achieved dynamics in a CFT$_1$ because we have coupled it to a large reservoir with which it can interact. However, we do not have dynamics in the normal sense in that the CFT$_1$ is not a self-contained deterministic theory: the equations of motion are nonlocal in time.

\section{A Line of Fixed Points} \label{sec:line}
Let us recall the SYK model. This is a model of $N\gg1$ Majorana fermions, with anticommutation relations $\{\chi_i, \chi_j\} = \delta_{i j}$ and action,  $S_{top} + S_{SYK}^{\rm int}$, where,
\be \label{Stop}
S_{top} =   \frac{1}{2} \sum_{i = 1}^N \int d\tau\, \chi_i \, \frac{d}{d \tau}\, \chi_i~
\ee
is the action for free Majorana fermions, and the interaction is, 
\be  \label{SSYK}
S_{SYK}^{\rm int} = \frac{(i)^{\frac{q}{2}}}{q!}\sum_{i_1, \ldots, i_q=1}^N\,\int d\tau \, J_{i_1\, i_2\, \ldots i_q}\,  \chi_{i_1}\chi_{i_2}\, \cdots \chi_{i_q}~,
\ee
where the coupling $J_{i_1, \ldots, i_q}$ is totally antisymmetric and, for each $i_1, \ldots, i_q$, is chosen from a Gaussian ensemble, with variance,
\be \label{disA}
\frac{1}{(q-1)!} \sum_{i_2, \ldots, i_q=1}^{N}\langle J_{i_1 i_2 \ldots i_q}  J_{i_1 i_2 \ldots i_q}\rangle= J^2~.
\ee
One can consider SYK for any even $q\geq 2$, with  $q=4$ being the prototypical case.

 In the UV, at zero coupling, the action is just (\ref{Stop}), and the fermions have a two-point function given by $\frac{1}{2} \sgn(\tau)$.  One of the central features of SYK is that the two-point function $G(\tau)$ of the fermions is  conformally invariant in the infrared. In particular, for $J |\tau| \gg 1$, one has, at leading order in $1/N$, 
\be
G(\tau) = b\frac{\sgn(\tau)}{|J\tau|^{2\Delta}}~,
\ee
where $b$ is given by,
\be \label{psiDelta}
\psi (\Delta) \equiv 2 i \cos (\pi \Delta) \Gamma(1-2\Delta) ~, \ \ \ \ \ b^{q} = - \frac{1}{\psi(\Delta) \psi(1-\Delta)} = \frac{1}{2\pi}(1 - 2\Delta)\tan\pi \Delta~,
\ee
and the IR dimension of the fermions is $\Delta = 1/q$. 

We would like to introduce a slight variant of SYK, conformal SYK (cSYK), that has $SL(2,R)$ invariance for all values of $J$. To do this we modify the UV part of the SYK action, replacing $S_{top}$ with an action $S_0$ that gives the fermions dimension $\Delta=1/q$ at the outset. In other words, we would like the UV two-point function to be, 
\be \label{G0}
 G_0(\tau) = \frac{1}{2}\frac{\sgn(\tau)}{|\tau|^{2\Delta}}~, \ \ \ \ \ \ \ \ \ \ \ \ \ G_0(\omega) = \frac{1}{2} \psi(\Delta) |\omega|^{2\Delta-1} \sgn(\omega)~.
\ee
In order to achieve this we replace $S_{top}$ with the bilocal action,
\be  \label{S0}
S_0 =-\frac{1}{2}\sum_{i = 1}^N \int d\tau_1 d\tau_2\, \chi_i(\tau_1)\, K(\tau_1- \tau_2)\, \chi_i(\tau_2)~,
\ee
where the kernel $K(\tau_1- \tau_2)$ is chosen to be the inverse of the propagator, 
\be \label{KS0}
K(\tau) = \int \frac{d\omega}{2 \pi }  e^{- i\omega \tau}\, G_0(\omega)^{-1} =- 2 b^q \frac{\sgn(\tau)}{|\tau|^{2(1-\Delta)}}~.
\ee
The action for cSYK is thus, 
\be \label{LACT}
S= S_0 + S_{SYK}^{\rm int}~.
\ee

One should note that unlike in SYK, here the coupling $J$ is dimensionless. Also, one can see that in the limit that $\Delta\rightarrow 0$ we get back the SYK action, as in this limit $K(\tau) \rightarrow \frac{d}{d \tau} \delta(\tau)$.

\subsection{Correlation Functions}

\subsection*{Two-Point Function}
\begin{figure}
\centering
\subfloat[]{
\includegraphics[width=2.5in]{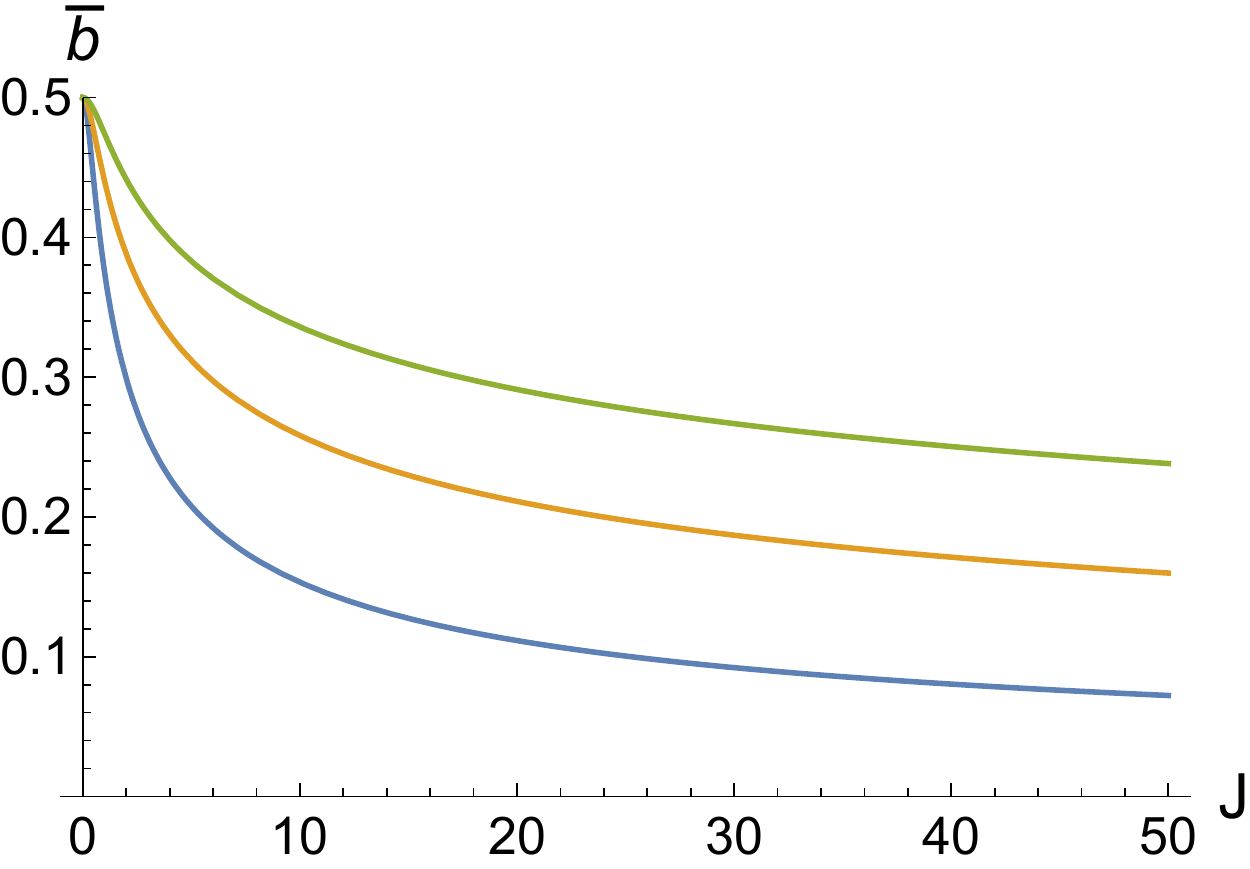}
} \ \ \ \ \ \ \ 
\subfloat[]{
\includegraphics[width=2.5in]{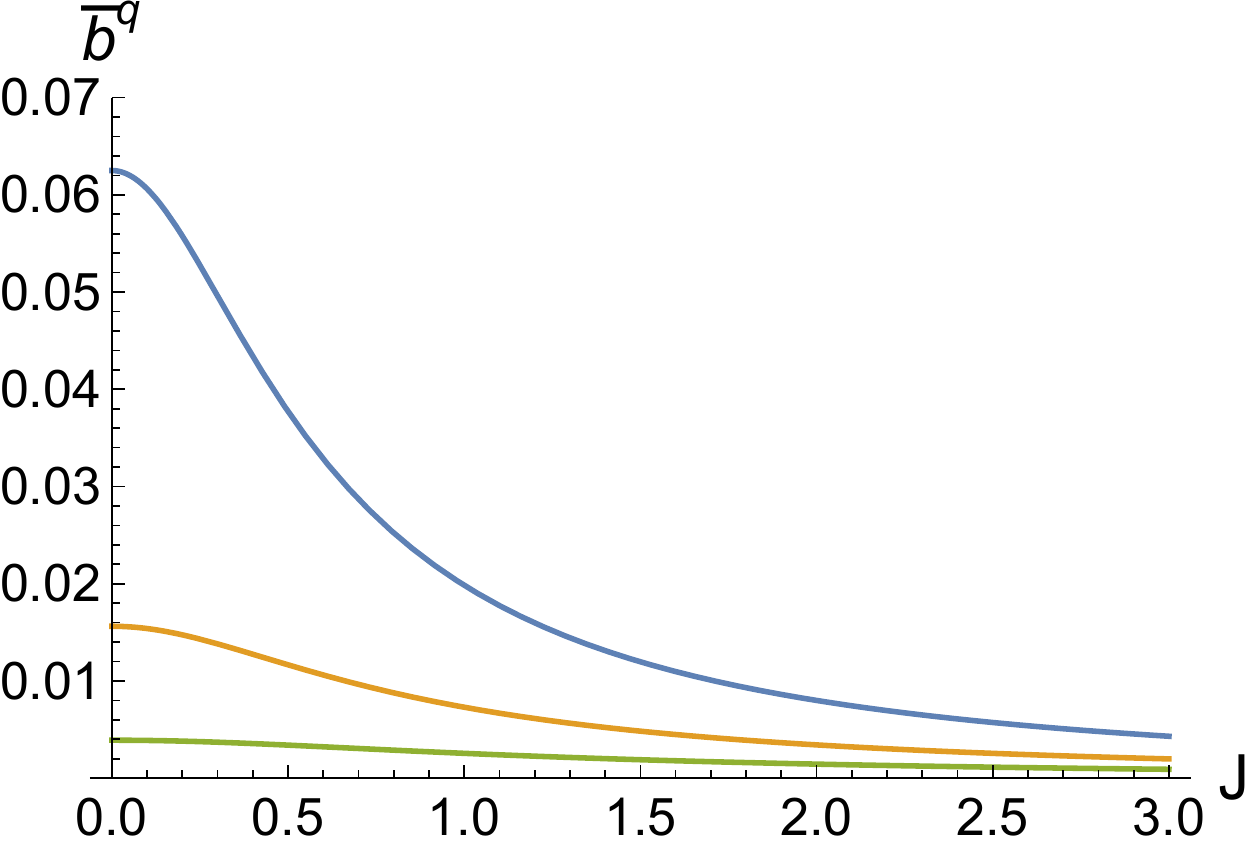}
}
\caption{(a) A plot of the running of the normalization of the cSYK fermion two-point function, ${\bar b}$,  as a function of $J$, for $q=8,6,4$. The larger the $q$, the slower the decay. (b) A plot of ${\bar b}^q$; this decreases with increasing $q$ (the opposite of ${\bar b}$). }\label{FigblPlot}
\end{figure}
Let us now look the Schwinger-Dyson equations for the two-point function of cSYK at large $N$. These are similar to SYK, the only change being that there is a different $G_0$, 
\bea \label{SD}
G(\omega)^{-1}&=& G_0(\omega)^{-1} - \Sigma(\omega)~, \\ \nonumber
\Sigma(\tau) &=& J^2 G(\tau)^{q-1}~.
\eea
In SYK, $G_0(\omega) = i/\omega$, and the equations have no known analytic solution away from the fixed points (except in the $q \to \infty$ limit.) For cSYK, $G_0(\omega)$ is given by (\ref{G0}) and one can immediately verify that (\ref{SD}) is solved by the conformal ansatz, 
\be \label{2pLine}
G(\tau) = {\bar b}\frac{\sgn(\tau)}{|\tau|^{2\Delta}}~,
\ee
where $\Delta = 1/q$ and the coefficient ${\bar b}$ satisfies the equation, 
\be \label{blEq}
\frac{{\bar b}^q}{1-  2{\bar b}} = \frac{1}{J^2 \psi(1-\Delta) \psi(\Delta)} =  \frac{1}{2\pi J^2} \(1 - 2\Delta\) \tan \pi \Delta= \frac{ b^q}{J^2}~.
\ee
At this level, the effect of varying the coupling is just to vary the prefactor ${\bar b}$. We can not analytically solve this equation, but we have plotted ${\bar b}$ in Fig.~\ref{FigblPlot}. For fixed $q$, in the limit that $J \ll 1 $ and $J\gg1$, we have, 
\bea \label{blw}
{\bar b} &=& \frac{1}{2} \( 1- \frac{ J^2}{(2 b)^q}+ \ldots\)~, \ \ \ \   \ \ \ \ \ \ \ \ \ \ \ J \ll 1~,  \\ \label{bls}
{\bar b} &=&\frac{b}{J^{2/q}} \( 1- \frac{2 b}{q J^{2/q}}+\ldots \)~, \ \ \ \ \  \ \ \ \ \ \ J \gg 1~.
\eea
At small $J$, we recover $G_0(\tau)$. At large $J$, we recover the SYK two-point function in the IR. This is reasonable; sufficiently deep in the infrared we forget about the UV. 

One can notice that the cSYK self-energy is proportional to the kernel in the action, 
\be
\Sigma(\tau) = \(1- \frac{1}{2 {\bar b}}\) K(\tau)~,
\ee
while the kernel is in turn related to the infrared SYK self-energy $\Sigma^{IR}_{SYK}$ through,
\be \label{KSYK}
K (\tau)=- \frac{ 2 b}{J^{2\Delta}} \Sigma^{IR}_{SYK}(\tau)~.
\ee

The theory we have discussed so far is at zero temperature.  At finite temperature, one must use a different action. In particular, the kernel (\ref{KS0}) appearing in $S_0$ should be replaced with, 
\be \label{Ktemp}
K(\tau) = - 2 b^{q} \frac{\sgn(\tau)}{\(\frac{\beta}{\pi} \sin \frac{\pi \tau}{\beta}\)^{2(1-\Delta)}}~,
\ee
and the time integration in the action should run from zero to $\beta$. One can check  that the solution of the Schwinger-Dyson equations is, 
\be
G(\tau) ={\bar b} \frac{\sgn(\tau)}{\(\frac{\beta}{\pi} \sin \frac{\pi \tau}{\beta}\)^{2 \Delta}}~.
\ee
Note that cSYK has SL(2,R) invariance, but not time reparameterization invariance. The action for free Majorana fermions has time reparameterization invariance, and the two-point function is $\frac{1}{2} \sgn(\tau)$ regardless of the temperature. In the infrared of SYK there is (almost) emergent time reparameterization invariance, which is spontaneously broken by the choice of the vacuum to $SL(2,R)$ invariance (which is explicitly broken at the level of the four-point function). However, at the level of the Schwinger-Dyson equations for the two-point function, one has for the IR of SYK, that if $G(\tau)$ is a solution then, 
\be 
\tau\rightarrow f(\tau)~, \ \ \ \ G(\tau_1, \tau_2) \rightarrow |f'(\tau_1)|^{\Delta} |f'(\tau)|^{\Delta} G(f(\tau_1), f(\tau_2))
\ee
is also a solution. 
Since we want to have a line of fixed points, we must explicitly break time reparameterization invariance through our choice of the UV action. With this in mind, we will however continue to use the zero-temperature theory. 

\subsection*{Variations and generalizations}
One can consider variants of SYK that do not have disorder. One way is to turn SYK into a tensor model \cite{Gurau:2009tw, Witten:2016iux, Klebanov:2016xxf}; to leading order in $1/N$, this gives the same results. The cSYK model can similarly be made into a tensor model.

 Another approach to removing disorder is to make the couplings $J_{i_1 \ldots i_q}$ be nearly static quantum variables \cite{MPRS}; at leading order in $1/N$ this also gives the same results for connected correlation functions. One way of doing this is to make the two-point function of the $J_{i_i \ldots i_q}$ have time dependance of the form $\langle J_{i_1 \ldots i_q}(\tau_1) J_{i_1 \ldots i_q}(\tau_2)\rangle = J^2 (q-1)! N^{-q+1}\, |\tau_{12}|^{-2\alpha}$, and then send  $\alpha$ to zero \cite{MS}. In fact, for cSYK, taking this as the two-point function of the $J_{i_1\ldots i_q}$, and choosing  
$\alpha = 1 - \Delta q$, we can get a two-parameter ``plane'' of fixed points. In particular, let $\Delta$ and $q$ be chosen independently, and take the action (\ref{LACT}), where $\Delta$ and $q$ are independent. There is now a line of fixed points,  where the two-point function is (\ref{2pLine}) and the magnitude ${\bar b}$ satisfies (\ref{blEq}). We have written these equations, as well as later equations for the OPE coefficients and three-point functions of bilinears, in a way so that they are valid for independent $\Delta$ and $q$ (in usual discussions of SYK, one has $\Delta = 1/q$, and $\Delta$ and $q$ are used interchangeably).~\footnote{In particular, Eqs.~\ref{KlK}, \ref{ghSYK}, \ref{cnSYK2}, \ref{cnmk1}, \ref{cnmk2}, are valid for any $\Delta, q$. Some of the other equations, in which we take various limits, such as (\ref{c0}), are restricted to $\Delta=1/q$.} We can try to make the action local, by sending $\Delta$ to zero. In order to have a solution to (\ref{blEq}), we must also send $J$ to zero, so that $J^2/\Delta$ remains constant. This turns out to be an uninteresting limit, in which the dimensions,  OPE coefficients, and three-point functions of all bilinear operators, with the exception of $\mO_0$, take on the $\Delta\rightarrow 0$ generalized free field answer.  

Finally, SYK has $O(N)$ symmetry after disorder averaging. One can add flavor, making the symmetry $O(N_1) \times O(N_2) \times\ldots \times O(N_f)$ \cite{GR}. The same can be done for cSYK.

\subsection*{UV finiteness}
Returning to SYK/ cSYK, the Schwinger-Dyson equations (\ref{SD}) can be rewritten as a single integral equation, 
\be \label{SD}
G (\tau_{12}) = G_0(\tau_{12}) + J^2 \int d\tau_a d\tau_b G_0(\tau_{1a})G(\tau_{ab})^{q-1} G(\tau_{b2})~.
\ee
For SYK, the way of solving this equation in the infrared is to drop the left-hand side of (\ref{SD}), and insert the conformal ansatz $G(\tau) = b\, \sgn(\tau) |\tau|^{-2/q}$ into (\ref{SD}) in order to find $b$. Naively,  it would appear that the integral is UV divergent. If this were the case, it would be an artifact of dropping the left-hand side of (\ref{SD}) as  SYK is super-renormalizable, and the two-point function behaves as  $\frac{1}{2} \sgn(\tau)$ in the UV. But, in fact,  even with the conformal ansatz for $G(\tau)$, the integral is  UV finite. The only potential divergence in the integral occurs for $\tau_a \rightarrow \tau_b$. To first order, near this region one has to integrate, 
\be 
\int d\tau_a d\tau_b\, \frac{\sgn(\tau_{ab})}{|\tau_{ab}|^{2(1 -\frac{1}{q})}}\( 1 + \# \tau_{ab} + \ldots\)~,
\ee
which is finite, as antisymmetry causes the leading term to vanish.  For cSYK, the integrals involved in summing the melon diagrams are again UV finite, as a result of antisymmetry. Unlike the super-renormalizable SYK,  cSYK is classically scale invariant and there could be quantum anomalies. The UV finiteness of the singlet sector of cSYK implies that the theory is indeed conformally invariant, at least to leading order in $1/N$, and perhaps to all orders in the $1/N$ expansion, for any value of $J$. The finiteness is apparently a consequence of the fermion antisymmetry and the randomness of the $J_{{i_1}\ldots {i_q}}$ couplings.

While the double integral in (\ref{SD}) is UV finite, it is not absolutely convergent. The standard way of evaluating these integrals in SYK has been to turn this convolution into a product of Fourier transforms. A more rigorous approach was recently discussed in \cite{Gurau}, and involves multiplying the frequency-space two-point function by  $e^{- \epsilon |\omega|}$. 

At the level of the four-point function, to leading order in $1/N$, one sums ladder diagrams. At order $J^2$ there are two ladder diagrams, with the second related to the first by exchange of $\tau_3 \leftrightarrow \tau_4$, 
\be
J^2 \int d\tau_a d\tau_b\,\[ G(\tau_{1a}) G(\tau_{a 3}) G(\tau_{2 b})G(\tau_{b 4}) G(\tau_{ab})^{q-2} - (\tau_3 \leftrightarrow \tau_4) \]~.
\ee
Either of these two terms individually has a UV divergence, from the region $\tau_a \rightarrow \tau_b$. However, the difference between the two terms is UV finite. Again, the fermion antisymmetry is responsible for  this. For similar reasons, all the ladder diagrams are UV finite. It is likely that any correlation function in cSYK, to leading nontrivial order in $1/N$, is UV finite. It is conceivable that sub-leading $1/N$ corrections could also be UV finite, but we have not checked this. 

\subsection*{Four-point function}
We now move on to discussing the fermion four-point function for cSYK. The primary $O(N)$ invariant  bilinear operators are, 
\be \label{O2n1}
\mO_{n} = \sum_{i=1}^N \sum_{k=0}^{2n+1} d_{n k}\, \partial_{\tau}^{k} \chi_i\, \partial_{\tau}^{2n+1 - k} \chi_i~,
\ee
where the coefficients $d_{n k}$ are chosen so that the operators are primary. 
The dimensions $h_n$ of the $\mO_n$ are found by solving the eigenvector equation,  in the notation of \cite{GR},
\be \label{EIG2}
{\bar g}(h)\, v(\tau_0; \tau_1, \tau_2) =  \int d\tau_3 d\tau_4\, K_l(\tau_1, \tau_2, \tau_3, \tau_4)\, v(\tau_0; \tau_3, \tau_4)~,
\ee
for the $h_n$ for which ${\bar g}(h_n) = 1$. The eigenvectors $ v(\tau_0; \tau_1, \tau_2) $ are conformal three-point functions $\langle \mO_n(\tau_0) \chi_i(\tau_1) \chi_i(\tau_2)\rangle$. 
The kernel $K_l$ is related to the SYK kernel by a simple factor, which in turn relates ${\bar g}(h)$ to the corresponding $g(h)$ in SYK,  
\be \label{KlK}
\frac{K_l}{K} = \Big(\frac{{\bar b}}{b}\Big)^q J^2~, \ \  \ \ \ \ \ \ \ \ {\bar g}(h) = \Big(\frac{{\bar b}}{b}\Big)^q J^2 g(h) = (1-2 {\bar b}) g(h)~,
\ee
where the eigenvalues $g(h)$ for SYK are,
\be \label{ghSYK}
g(h) = - (q-1) \frac{\psi(\Delta)}{\psi(1-\Delta)}  \frac{\psi(1- \Delta - \frac{h}{2})}{\psi(\Delta - \frac{h}{2})}~.
\ee
The equation for the dimensions $h_n$, ${\bar g}(h_n) = 1$, can only be solved numerically; for general $q$ and general $J$, the dimensions $h_n$ have an order-one shift from $2 \Delta + 2n+1$, which goes to zero for large $n$. It is instructive to see how the dimensions change as the coupling $J$ is varied. 
At weak coupling, $1-2 {\bar b} \rightarrow 0$, and so in order to have ${\bar g}(h)=1$ one must have $g(h)$ diverge, which means $h$ approaches the free value, 
\be
h \rightarrow 2 \Delta + 2n +1, \ \ \ \ \ \text{as  } \ \ J\rightarrow 0~.
\ee
On the other hand, at strong coupling, $ {\bar b} \rightarrow 0$, and so the dimensions $h_n$ approach those of the infrared of SYK, given by the solutions  to $g(h_n) = 1$.

An interesting operator is the lowest dimension operator, $\mO_0$. It has dimension $h_0$ that increases from $1$ at weak coupling to $2$ at strong coupling. 
For strong coupling its dimension is, 
\be
h_0  =2 - \varepsilon_0~, \ \ \ \ \ J \gg 1~,
\ee
where,
\be
\varepsilon_0 =  2 {\bar b} \[\frac{\pi}{\sin(2\pi/q)} - \frac{q(6+q(q-6))}{2(q-2)(q-1)}\]^{-1}~.
\ee
Recalling the scaling of ${\bar b}$ at strong coupling (\ref{bls}), we see that $\epsilon_0$ scales as $J^{-2\Delta}$ for strong coupling.

Next, let us look at the OPE coefficients $c_n$, 
\be \label{OPE}
\frac{1}{N} \sum_{i=1}^N \chi_i(\tau_1) \chi_i (\tau_2) = \frac{1}{\sqrt{N}}\sum_{n=1}^{\infty} c_n\, G(\tau_{12}) |\tau_{12}|^{h_n} ( 1+ \frac{1}{2} \tau_{12} \partial_{2} + \ldots)\mO_n(\tau_2)~.
\ee
The result for the $c_n$ for cSYK is a simple extension of the SYK answer \cite{MS}, giving, 
\be \label{cnSYK2}
\!\! c_n^2\! =\!  \alpha_0(q, \Delta)\frac{(h_n - 1/2)}{\pi \tan(\pi h_n/2)} \frac{\Gamma(h_n)^2}{\Gamma(2 h_n)} \frac{1}{(1-2 {\bar b})^2 g'(h_n)} ~,  \ \ \ \text{where}\ \ \alpha_0(q, \Delta)\! =\! \frac{2\pi }{(q-1) (1-2\Delta) \tan \pi \Delta}~.
\ee
The OPE coefficients for SYK are the above, but without the factor of $(1-2 {\bar b})^2$ in the denominator. Note that the $h_n$ entering here are the dimensions of the $\mO_n$, found previously by solving ${\bar g}(h_n) = 1$. These coefficients are valid for all $J$; the $J$ dependence is encoded explicitly in the $1-2 {\bar b}$ factor, as well as implicitly in the dimensions $h_n$. 

Let us look in particular at the behavior of $c_0^2$ at strong coupling. For SYK, it was the $\mO_0$ operator that broke conformal invariance, so for cSYK we expect that $c_0^2$ will diverge as $J$ goes to infinity, since in this limit we are approaching the IR of SYK. Indeed see from (\ref{cnSYK2}) that there is a divergence, which comes entirely from the  $\tan(\pi h_0/2)$ factor in the denominator. In particular, to leading order, 
\be \label{c0}
c_0^2 = \frac{C_0^2}{\varepsilon_0}~, \ \ \ \ \ \ J \gg 1~, 
\ee
where, 
\be
C_0^2 =  \frac{2 q}{\pi \tan(\pi/q)} \[ \frac{2\pi (q-2)(q-1)}{\sin 2\pi/q} - q(6+q(q-6))\]^{-1}~.
\ee
The four-point function of the fermions is given by a sum of conformal blocks of the  $\mO_n$ operators. For cSYK, unlike for SYK, the $\mO_0$ operator is on the same footing as the other operators and the four-point function has $SL(2,R)$ invariance. 

\subsection*{Three-point function of bilinears}
As with the fermion four-point function, it is trivial to extend the infrared SYK results for the bilinear three-point function to the cSYK model. This is done in Appendix.~\ref{ap:three}.

The coefficient of the three-point function of the $\mO_n$, $\langle \mO_n \mO_m \mO_k\rangle$,  translates into the coefficient $\lambda_{ n m k}$ of the cubic couplings of the bulk fields $\phi_n$ dual to $\mO_n$. The bulk Lagrangian, up to cubic order, is
\be \label{sb1}
S_{bulk} = \int d^2 x \sqrt{g} \[ \sum_{n=0}^{\infty} \frac{1}{2}\( (\partial \phi_n^2) + m_n^2 \phi_n^2 \)+ \frac{1}{\sqrt{N}} \sum_{n, m, k=0}^{\infty} \lambda_{n m k } \phi_n \phi_m \phi_k+ O(\frac{1}{N})\]~.
\ee
This is for cSYK; for the infrared of SYK the sums start at $n=1$ rather than $n=0$. Let us consider the limit of large $J$. Then $\lambda_{n m k}$, for $n, m, k \neq 0$, approach the infrared SYK values, which are of course independent of $J$. The novel part of (\ref{sb1}), relative to the infrared of SYK, are the terms involving $\phi_0$. 
 Let us write out the form of the piece of the bulk Lagrangian $S_{\phi_0} \subset S_{bulk}$ that involves $\phi_0$, up to cubic order and at large $J$,
\be \label{Sphi0}
\!\!\! S_{\phi_0}\! =\!\!  \int\!\! d^2 x \sqrt{g}\! \[\!  \frac{1}{2}\( (\partial \phi_0^2) + m_0^2 \phi_0^2 \)\! +\! \frac{1}{\sqrt{N}} (J^{\frac{3}{q}} \bar{\lambda}_{0 0 0 } \phi_0^3 + J^{\frac{2}{q}} \sum_{n=1}^{\infty} \bar{\lambda}_{0 0 n} \phi_0^2 \phi_n + J^{\frac{1}{q}}\!\!\! \sum_{n, m=1}^{\infty} \bar{\lambda}_{0 n m} \phi_0 \phi_n \phi_m)\]~,
\ee
where we have explicitly separated out the $J$ dependence of the coupling, so that $\bar{\lambda}_{n m k}$ is independent of $J$. The divergence of these couplings as $J\rightarrow \infty$ is simply a consequence of the divergence of the OPE coefficient for $\mO_0$ from 2 fermions, (\ref{c0}).~\footnote{In particular, $\lambda_{n m k}$ follows from the coefficient $c_{n m k}= c_{n mk}^{(1)} + c_{n m k}^{(2)}$ of the bilinear three-point function, (\ref{OOO}). For $c_{n m k}^{(1)}$, given by (\ref{cnmk1}), there is a contribution from the OPE coefficients, as well as $\mI_{n m k}^{(1)}$ (\ref{I1exact}). One can verify that $\mI_{n m k}^{(1)}$ is finite if any or all of the $n,m, k$ are zero, and the coupling is strong (so that $h_0$ approaches two). In fact, $\mI_{0 00}^{(1)}$ can be seen to vanish at strong coupling, as $\mI_{0 0 0 }^{(1)} \sim \varepsilon$. The other piece, $c_{n m k}^{(2)}$, aside from the OPE coefficients, involves $\mI_{n m k}^{(2)}$ which is determined by the integral (\ref{I2}), which is manifestly finite for any or all of the $h=2$. Note that in writing (\ref{Sphi0}) we have kept $q$ finite, while taking the $J\rightarrow \infty$ limit. } In SYK, the analogue of (\ref{c0}), forced us to move slightly away from the infrared, leading to the breaking of $SL(2,R)$ invariance, and in the bulk, to large backreaction. For cSYK, we see that $J\rightarrow \infty$ leads to the divergence of the cubic couplings, so one could say that we should move slightly away from this limit, though here, since there is a line of fixed points, large and finite $J$ is no more involved than infinite $J$.

\section{Discussion}\label{sec:bulk}
In usual studies of AdS/CFT, the bulk gravitational sector is taken to be Einstein gravity, at least to leading order. In AdS$_2$ this is not possible: variation of the Einstein-Hilbert action, $ \int d^{2} x \sqrt{g} (R+2)$, gives identically zero. One option is to simply not have an Einstein-Hilbert term, and not vary the metric. This is not something one would normally consider, because a duality with a nongravitational bulk is not really holography. In the context of AdS$_2$, it is more reasonable, since there is no normal gravity anyway. If one does explore this option, such as for cSYK, then the dual CFT$_1$ should have as many degrees of freedom as a 2d theory.  For the cSYK  model, this is achieved as a result of the  bilocal term in the action, which makes it like a subsector of a 2d theory. 
Another option is to invent an action  to serve as a toy model for a gravity-like theory in two-dimensions. Jackiw-Teitelboim gravity \cite{Jackiw:1984je, Teitelboim:1983ux} is of this type. A dilaton field $\phi$ acts as a Lagrange multiplier, so that variation of the action $ \int d^2 x \sqrt{g}\, \phi\[\frac{1}{16 \pi G} (R+2) + \mathcal{L}_{m}\]$ gives an equation that resembles Einstein's equations. In fact, Jackiw-Teitelboim  is a good model of gravity, since it naturally arises from  dimensional reduction of  Einstein gravity in higher dimensions. 

The bulk dual of the infrared of SYK has Jackiw-Teitelboim gravity in the bulk. More specifically, the bulk Lagrangian, to leading order at strong coupling, is, 
\be \label{JT}
-\frac{1}{16 \pi G}\int d^2 x\sqrt{g}\, \phi \(R+2\) + \int d^2 x \sqrt{g}\, \mathcal{L}_{matter}~,
\ee
where the cosmological constant is $-2$ in units in which the AdS radius is one, and we have left out a topological term and the boundary term, and $\mathcal{L}_{matter}$ is the matter Lagrangian involving the tower of fields $\phi_n$ for $n\geq 1$. Here the matter is not coupled to the dilaton, so the metric can just be fixed to be pure AdS. The equations of motion relate the dilaton to the matter stress-tensor. The dilaton does not act like a normal scalar in AdS; it is not dual to an operator, and it grows near the boundary, $z=0$. As recognized in \cite{AP}, if one considers excited states then, since the value of $\phi$ changes even near  $z=0$, one must adjust the shape of the boundary curve on which the CFT$_1$ lives, in order to maintain constant $\phi$ on it. This ``backreaction'' leads to  breaking of conformal invariance in the CFT four-point function. In particular, the piece of the CFT four-point function that breaks conformal invariance, as computed from the bulk using (\ref{JT}), is the same as the piece of the SYK four-point function that breaks conformal invariance \cite{AP, MS, MSY, Jensen:2016pah}.

Unlike  for SYK, the bulk dual of cSYK  has no dilaton and consists of field theory on a fixed background. The bulk Lagrangian is,
\be
\int d^2 x \sqrt{g}\( \mathcal{L}_{\phi_0} + \mathcal{L}_{matter}'\)~,
\ee
where $\mathcal{L}_{\phi_0}$ consists of terms containing the scalar $\phi_0$, dual to $\mO_0 = \chi_i \partial_{\tau} \chi_i$, while $\mathcal{L}_{matter}'$ consists of  terms  containing exclusively the rest of the fields, $\phi_n$ with $n\geq 1$. In the limit of strong coupling, $J \gg 1$, one has $\mathcal{L}_{matter}' = \mathcal{L}_{matter}$. Since cSYK has a line of fixed points, one can consider the bulk dual for any $J$. In the appendix we compute the cubic couplings at all values of $J$, finding that  $\lambda_{n m k}$ become independent of  $J$ at large $n, m,k$. It will be interesting to see how the form of the bulk quartic couplings depends on the coupling $J$ \cite{GRprog}.

\bigskip

\section*{Acknowledgements} \noindent We thank R.~Gurau and A.~Tseytlin for helpful discussions.  
This work was supported by NSF grant 1125915. VR thanks the Aspen Center for Physics, NSF Grant PHY-1066293, for hospitality while this work was being completed. 
\appendix{}

\section{OPE and Three-Point Function of Bilinears}  \label{ap:three}
In this appendix we  extend to cSYK the results of \cite{GR2} for the three-point function of $O(N)$ invariant  fermion bilinear singlets in the infrared of SYK. We study how    these three-point functions in cSYK change as one varies the coupling $J$ from weak  to strong. 
\subsection{OPE}
We begin by considering the weak and strong coupling limits of the OPE of two fermions in cSYK, as given by (\ref{cnSYK2}).
\subsubsection*{OPE for weak coupling}
Let us look more closely at the behavior of the dimensions $h_n$ of the bilinear singlets $\mO_n$ for cSYK for weak coupling. In this case, $h_n \approx 2 \Delta + 2n +1$, and near these $h_n$ we can expand $g(h)$ (\ref{ghSYK})  as, 
\be \label{ghpole}
g(h) = \frac{\gamma_n}{ h - (2n+2\Delta+1)}+ \ldots~, \ \ \ \ \ \ \ \gamma_n = \frac{4 (q-1)(q-2)}{q^2} \frac{\Gamma(2n+\frac{4}{q})}{\Gamma(2n+2)\Gamma(\frac{2}{q}) \Gamma(1+\frac{2}{q})}~.
\ee
To find the $h_n$ to leading order in $J^2$, we set ${\bar g}(h_n) = 1$. Recalling (\ref{KlK}), this gives, 
\be \label{hnWeak}
h_n = 2\Delta+ 2n+1 + (1-2 {\bar b}) \gamma_n ~.
\ee
Recall from (\ref{blw}) that $(1-2{\bar b})$ scales like $J^2$ for small $J$. 
Furthermore, from differentiating  (\ref{ghpole}) we get that $(1-2{\bar b})^2 g'(h_n) = -1/\gamma_n$, and hence the OPE coefficients (\ref{cnSYK2}), in the limit of weak coupling $J$, are, 
\be
c_n^2 = - \alpha_0(q)\frac{(h_n - 1/2)}{\pi \tan(\pi h_n/2)} \frac{\Gamma(h_n)^2}{\Gamma(2 h_n)}\gamma_n~, \ \ \ \ J\ll 1~.
\ee
We can now take the limit of $J=0$, to find, 
\be
c_n^2 = \frac{2}{q\, \Gamma(\frac{2}{q})^2} \frac{ (4+q + 4n q)\Gamma( 2n +1 +\frac{2}{q})^2 \, \Gamma(2n + \frac{4}{q})}{\Gamma(2n+2) \Gamma(4n + 2 + \frac{4}{q}) }~, \ \ \ \ \ J=0~.
\ee
In the limit that $q\rightarrow \infty$ this simplifies to,
\be \label{cnSmall}
c_n^2 = \frac{1}{q^2} \frac{2}{(2n+1)}\frac{\sqrt{\pi} \Gamma(2n)}{\Gamma(2n+ \frac{1}{2}) 2^{4n-2}}~, \ \ \  J=0~, \ \ q\rightarrow \infty~.
\ee

\subsubsection*{OPE for strong coupling and large $q$}
Now let us look at the $h_n$ and the $c_n^2$ for cSYK at strong coupling. At strong coupling, these have the same behavior for cSYK as for SYK. For SYK, these do not have a simple analytic form for general $q$, but they are simple for large $q$. In particular, for large $q$, we have for the dimensions \cite{GR2}, 
\be \label{hnq}
h_n = 2 n + 1 + 2 \epsilon_{n}~, \ \ \ \  \epsilon_{n} = \frac{1}{q} \frac{2 n^2 +n + 1}{2 n^2 +n - 1}~, \  \ \ \ \ n\geq1~,\ \ \ q\gg 1~,\ \\  \ \ \ \ J\rightarrow \infty~,
\ee
while the OPE coefficients are,
\be \label{cnLarge}
c_n^2 = \eps_n^2 \frac{n(1+2n)}{\(n(1+2n)+1\)\(n(1+2n)-1\)}\frac{\sqrt{\pi} \Gamma(2n+1)}{\Gamma(2n+\frac{1}{2}) 2^{4n-2}}~, \  \ \ \ \ n\geq1~, \ \ \ \ \ q \gg 1~~, \ \ \ J\rightarrow \infty~.
\ee

It is interesting to look at the ratio of the OPE coefficients at strong coupling (\ref{cnLarge}) to those at weak coupling (\ref{cnSmall}), for large $q$, 
\be \label{cnRatio}
\frac{c_n^2 (J=\infty)}{c_n^2(J=0)} = \frac{n^2 (2n+1)^2 (n(2n+1)+1)}{(n(2n+1) - 1)^3}~, \ \ \ \ \ q\rightarrow \infty~.
\ee

\subsection{Three-point Function Bilinears}
The three-point function of the fermion bilinears $\mO_n$ (\ref{O2n1}) for $n\geq 1$ for the infrared of SYK was computed in \cite{GR2}, to leading nontrivial order in $1/N$. By conformal invariance, it is fixed to take the form,
\be \label{OOO}
\langle \mO_{n}(\tau_1) \mO_{m}(\tau_2) \mO_{k}(\tau_3)\rangle =  \frac{1}{\sqrt{N}}\frac{ c_{n m k}}{|\tau_{12}|^{h_{n} +h_{m}  - h_{k}}| \tau_{23}|^{h_{m} + h_{k} - h_{n}} |\tau_{31}|^{h_{k} + h_{n} - h_{m}}}~.
\ee
There were two classes of Feynman diagrams that contributed to the fermion six-point function, out of which the bilinear three-point function was extracted. It was therefore useful to split the coefficient $c_{n m k}$ into two terms,
\be
c_{n m k} = c_{n m k}^{(1)} + c_{n m k}^{(2)}~, 
\ee
with the first denoting the contribution coming from diagrams that were called ``contact'' diagrams, and the second coming from planar diagrams. 

The first piece, $c_{n m k}^{(1)}$ is given by, 
\be \label{cnmk1}
c_{n m k}^{(1)} = (1- 2 {\bar b}) c_n c_m c_k\,  b^q (q-1)(q-2)\, \mI_{n m k}^{(1)}~,
\ee
where
\be \label{I1exact}
\!\!\mathcal{I}_{n m k}^{(1)}\! =\!\frac{  \sqrt{\pi}\, 2^{h_n + h_m + h_k -1}\, \Gamma(1\!-\!h_n) \Gamma(1\!-\!h_m) \Gamma(1\!-\!h_k)}{\Gamma\(\frac{3 - h_n - h_m - h_k}{2}\)}\! \[\rho(h_n, h_m, h_k)\! +\! \rho(h_m, h_k, h_n)\!+\! \rho(h_k, h_n, h_m) \]~,
\ee
where $c_n$ are the OPE coefficients (\ref{cnSYK2}), $b$ is given by (\ref{psiDelta}), ${\bar b}$ is given in (\ref{blEq}), and, 
\be
\rho(h_n, h_m, h_k) = \frac{\Gamma(\frac{h_m +h_k - h_n}{2})}{\Gamma(\frac{2-h_n-h_m +h_k}{2})\Gamma(\frac{2-h_n-h_k+h_m}{2})}\( 1+ \frac{\sin(\pi h_m)}{\sin(\pi h_k) - \sin(\pi h_n + \pi h_m)}\)~.
\ee
Here we have generalized the infrared of SYK result to cSYK. The only change is that we picked up a factor of $J^2 ({\bar b}/b)^q = 1-2 {\bar b}$ to account for the different normalization of the two-point function. In particular, the change in normalization of the two-point function $G$ is $b J^{-2\Delta} \rightarrow {\bar b}$, and there was a factor of $G^q$ that entered into getting this result.

The other piece, $c_{n n k }^{(2)}$,  is more involved, and is given by,
\be \label{cnmk2}
c_{ n m k}^{(2)} = c_n c_m c_k\,  \xi_n \xi_m \xi_k\, \mI_{n m k}^{(2)}~.
\ee
where $c_n$ are the OPE coefficients (\ref{cnSYK2}), while the factor $\xi_n$ is,
\be \label{eq:xi}
\xi_n =  b^q\, \pi^{\frac{1}{2}} \frac{\Gamma(1 - \Delta + \frac{h_n}{2})}{\Gamma(\frac{1}{2} + \Delta - \frac{h_n}{2})}\frac{\Gamma(\frac{1}{2} - \frac{h_n}{2})}{\Gamma(\frac{h_n}{2})}   \frac{\Gamma\(\Delta\)}{\Gamma\( \frac{3}{2} - \Delta\)}~,
\ee
and $\mI_{n m k }^{(2)}$ is the coefficient coming from the integral, 
\be \label{I2}
\!\!\!\!\!\!\!\!\! I_{n m k}^{(2)} \!= \!\!\! \int\!\! d\tau_a d\tau_b d\tau_c \frac{  -\sgn(\tau_{1a} \tau_{1b} \tau_{2a} \tau_{2c} \tau_{3b} \tau_{3c})  |\tau_{ab}|^{h_n  \!- \!1} |\tau_{ca}|^{h_m \!- \!1} |\tau_{b c}|^{h_k  \!- \!1}}{|\tau_{1a}|^{h_n \!- \!1+2\Delta} |\tau_{1b}|^{h_n +1-2\Delta} |\tau_{2 c}|^{h_m  \!- 1\!+2\Delta} |\tau_{2a}|^{h_m+1-2\Delta} |\tau_{3 b}|^{h_k  \!-1 \!+2\Delta} |\tau_{3 c}|^{h_k +1-2\Delta}}, 
\ee
which, since it transforms as a conformal three-point function, is characterized by the number $\mI_{n m k }^{(2)}$,
\be \label{6intA}
I_{n m k}^{(2)} (\tau_1, \tau_2, \tau_3) =\frac{\mI^{(2)}_{n m  k } }{ |\tau_{12}|^{h_n+h_m-h_k}  |\tau_{13}|^{h_n +h_k - h_m} |\tau_{23}|^{h_m+h_k -h_n}}~.
\ee
The quoted result for $c_{ n m k}^{(2)}$, while derived for the infrared of SYK,  is also valid for cSYK, without any explicit changes. There are of course implicit differences, in  that the OPE coefficients $c_n$ and the dimensions $h_n$ are now functions of the coupling $J$. 

\subsubsection*{Large $q$, strong coupling}
In the limit of strong coupling, the $c_n$ and $h_n$ of cSYK approach those of the infrared of SYK, and the same holds for $c_{n m k}$. While $c_{n m k}^{(1)}$ has an explicit form for any $q$, in \cite{GR2} it was possible to evaluate the integral $I_{n m k}^{(2)}$ entering $c_{n m k}^{(2)}$ only in the case of large $q$. In particular, at large $q$, one has for $\mI_{n m k}^{(2)}$,
\be \label{Cscri2}
\mI_{n m k}^{(2)} = s_{n m k}^{(2)} 
\( 2 \frac{\eps_n^+ + \eps_m^-}{\eps_n^+ \eps_m^-}\frac{\eps_m^+ + \eps_k^-}{\eps_m^+ \eps_k^-}\frac{\eps_k^+ + \eps_n^-}{\eps_k^+ \eps_n^-} - \frac{1}{\eps_n^+\eps_m^+\eps_k^+} - \frac{1}{\eps_n^- \eps_m^-\eps_k^-}\)
~, \ \ \ \ \ \ \ q\gg1,\ \ \  J\gg1~,
\ee
where $\eps_n^{\pm} \equiv \eps_n \pm \Delta$, where $\eps_n$ is given in (\ref{hnq}), and $s_{n m k}^{(2)} $ is a finite triple sum, given by Eq.~3.40 of \cite{GR2}, which we will not quote here. The appearance of $\eps_n$ here is a result of the dimensions $h_n$ of the bilinear having a deviation from $2n+1$ that is given by $2 \eps_n$, see Eq.~\ref{hnq}. Also, in the large $q$ limit, the factor $\xi_n$ (\ref{eq:xi}) simplifies to, 
\be
\xi_n = \(n + \frac{1}{2}\) \( 1- \frac{1}{\eps_n q}\) ~, \ \ \ \ \ \ q\gg1,\ \ \  J\gg1~,
\ee
while the $c_n$ in the large $q$ limit were given in (\ref{cnLarge}). 

\subsubsection*{Large $q$, weak coupling}
Now let us look at $c_{n m k}$ in the limit of weak coupling, $J\rightarrow0$. Due to the factor of $1-2{\bar b}$ in (\ref{cnmk1}), the piece $c_{n m k}^{(1)}$ goes to zero in this limit. This is as it should be, because at $J=0$ there are no contact diagrams. For the other piece, $c_{n m k}^{(2)}$, we need to evaluate the integral (\ref{I2}). Let us work at large $q$. Then we can immediatly establish the weak coupling result from the strong coupling one, (\ref{Cscri2}). The only change that we need to make is to account for the difference in deviation from $2n+1$ of the dimensions $h_n$. In particular, comparing (\ref{hnq}) and (\ref{hnWeak}) we see that what was $2\eps_n$ should be replaced with $2\Delta+ (1-2 {\bar b})\gamma_n$ (for large $q$ and weak coupling, this factor is much less than one). With this replacement, 
\be
\mI_{n m k}^{(2)}  = s_{n m k}^{(2)} \frac{8}{(1-2 {\bar b})^3}\, \frac{1}{\gamma_n \gamma_m \gamma_k}~, \ \ \ \ q\gg1,\ \ \  J\rightarrow0~.
\ee
Also, we have that $\xi_n$ simplifies to, 
\be
\xi_n =\(n + \frac{1}{2}\) \( 1- \frac{1}{1 + \frac{(1-2{\bar b}) \gamma_n q}{2}}\) \approx\frac{1}{2} \(n + \frac{1}{2}\) (1-2 {\bar b}) \gamma_n q~, \ \ \ \ \ \ q\gg1,\ \ \  J\rightarrow 0~.
\ee
Combining all the pieces as indicated by (\ref{cnmk2}), where $c_n$ was given in (\ref{cnSmall}),  we get, 
\be \label{cnmkFree}
c_{n m k}^{(2)} =   \frac{8}{N_n^{free} N_m^{free} N_k^{free} }\, s_{n m k}^{(2)}~, \ \ \ \ \ \ q\gg1,\ \ \  J= 0~,
\ee 
where
\be
(N_n^{free})^2 =  \frac{2^{4n+1}}{(2n+1)  }\frac{\Gamma(2n+\frac{1}{2})}{\sqrt{\pi}\,  \Gamma(2n)}~.
\ee
The result (\ref{cnmkFree}) of course matches what one  finds by computing the three-point function through Wick contractions, Eq.~A.12 of \cite{GR2}. 

\subsubsection*{Large $q$, from weak to strong coupling}
In \cite{GR2} we computed the three-point function of the bilinears for SYK in the infrared. This contained two pieces, $c_{n m k}^{(1)}$ and $c_{n m k}^{(2)}$, coming from a sum over Feynman diagrams which we called ``contact'' and planar, respectively. To compute $c_{n m k}^{(2)}$ required evaluating the  integral (\ref{I2}), which we were able to do at large $q$. The result involved a nontrivial triple sum $s_{n m k}^{(2)}$.  As a point of comparison, in \cite{GR2} we also computed the three-point function of bilinears in a generalized free field theory. This was a simple computation, just given by Wick contractions, and, surprisingly, the answer was related in a simple way to $c_{n m k}^{(2)}$, again involving the sum  $s_{n m k}^{(2)}$. It was somewhat mysterious why the results are so similar. 

Here, we have worked with cSYK, which has a line of fixed points, reducing to the generalized free field theory as $J\rightarrow 0$, and giving the infrared of SYK for $J\rightarrow \infty$. One advantage is that we have a single unified expression for the three-point function of bilinears, for any $J$. We can, in particular, look at it in the limit of small $J$, and get (\ref{cnmkFree}), thereby recovering the generalized free field answer obtained previously by Wick contractions.  In doing this, one can say that we have taken the four-point  function for the generalized free field theory, and written it in an especially complicated way, as a sum of conformal blocks, and then proceeded to use this to find the fermion six-point function, and hence the three-point function of bilinears. 

This exercise explains the similarity between the contribution of the planar diagrams for the three-point function of bilinears in the infrared of SYK at large $q$, and the three-point function of bilinears in a generalized free field theory. Namely, at large $q$ the dimensions $h_n$ of the bilinears approach the free dimensions, $2n+1$, and are characterized by the deviation $\eps_n$ (\ref{hnq}) from this. The $\eps_n$, combined with the ratio of OPE coefficients between those for the infrared of SYK and those for the generalized free theory, (\ref{cnRatio}), fully characterize the differences of the result.

\vspace{.5cm}

\bibliographystyle{utphys}

\end{document}